# Polariton properties in bigyrotropic medium

**I. V. Dzedolik, O. S. Karakchieva**

Taurida National V. I. Vernadsky University, 4, Vernadsky Av., Simferopol, 95007, Ukraine
dzedolik@crimea.edu

The spectra of polaritons in bigyrotropic medium are received. The ion and electron polarizations of medium under influence of high-frequency electromagnetic and external magnetostatic field are considered. It is shown that by varying of the external magnetostatic field one can control polariton spectrum and velocity of polaritons.

**Key words:** bigyrotropic medium , spectra of polaritons, external magnetostatic field, polariton velosity.

## 1. Introduction

It is known that the polaritons in the dielectric medium are the collective excitations (quasiparticles) corresponding to the bound states of photons and phonons [1-3]. The polaritons in bigyrotropic medium, i.e. in the dielectric medium with a magnetic subsystem, represent the quasiparticles arising at interaction of photons, phonons and magnons in such frequency range where permittivity and permeability of medium are not equal to unity, and the both of them depend on the frequency of electromagnetic field. The polaritons are subdivided to optical ones arising in nonmagnetic medium, and magnetic ones, which arise in a medium with magnetic subsystem [1, 4, 5]. In the bigyrotropic medium the polaritons show the dielectric and magnetic properties in the proper frequency ranges. In this case the polaritons have a combination of properties of both types [5, 6].

The polariton properties still attract attention of researchers because the interaction of electromagnetic field with dielectric medium in particular containing magnetic subsystem can be more adequately explained by polariton dynamics [7-14]. Collective excitations in a medium can be described by means of the microscopic multiparticle and macroscopic classical approaches which give the same results. In this paper we are using the macroscopic classical approach.

The external magnetostatic field superposed on bigyrotropic medium changes the both permittivity and permeability and thus influences on the polariton spectrum. It is possible to control the parameters of polaritons, in particular velocities of polaritons by changing the direction and intensity of the magnetostatic field. In this paper we compare the polariton spectra at perpendicular and parallel directions of magnetostatic field relatively to the wavevector and explore the influence of magnetostatic field on velocity of polaritons.

**2. Theoretical model**

Let us consider the elementary macroscopic model of polariton generation in the dielectric medium like ionic crystal with magnetic subsystem. In macroscopic model the dipole response of isotropic crystal of cubic system can be described by the equations of ion motion and outer shell electron motion [11, 14]. The equation of ion motion in a unit cell of crystal lattice looks like

$$\frac{d^2\mathbf{R}}{dt^2} + \Gamma \frac{d\mathbf{R}}{dt} - \frac{d\mathbf{R}}{dt} \times \boldsymbol{\omega}_{IB} + \Omega_\perp^2 \mathbf{R} = \frac{e_{eff}}{m_{eff}} \mathbf{E}, \qquad (1)$$

where $\mathbf{R} = \mathbf{R}_+ - \mathbf{R}_-$ is the radius-vector describing the displacement of positive and negative ions, $\Omega_\perp = \sqrt{q_1/m_{eff}}$ is the eigenfrequency of elastic lattice oscillations (frequency of transverse phonons), $\boldsymbol{\omega}_{IB} = e_{eff}\mathbf{B}_0/m_{eff}c$ is the Larmor frequency of oscillations of lattice charges in the external magnetostatic field, $m_{eff}$ is the effective mass, $e_{eff}$ is the effective charge of a unit cell, $\Gamma$ is the damping factor. The equation of electron motion is

$$\frac{d^2\mathbf{r}}{dt^2} + \Gamma \frac{d\mathbf{r}}{dt} + \frac{d\mathbf{r}}{dt} \times \boldsymbol{\omega}_{eB} + \omega_0^2 \mathbf{r} = -\frac{e}{m} \mathbf{E}, \qquad (2)$$

where $\omega_0$ is the resonance electron frequency, $e$ and $m$ are the charge and mass of electron, $\boldsymbol{\omega}_{eB} = e\mathbf{B}_0/mc$.

The tensor of medium permittivity $\varepsilon_{ij}$ is defined by the expression $D_i = \varepsilon_{ij}E_j = E_i + 4\pi P_i$, $i, j = x, y, z$, for electric displacement vector, where $P_i = e_{eff}N_C R_i - eN_e r_i$ is the component of medium polarization vector, $N_C$ and $N_e$ are the number of cells and number of electrons in unit volume. We obtain the expressions for radius-vectors $R_{x,y,z}$, $r_{x,y,z}$ by solving the equations (1) and (2). The components of tensor $\varepsilon_{ij}$ [11] are

$$\varepsilon_{xx} = 1 + \frac{\omega_{Pi}^2}{\Delta_{IB}}\left(\tilde{\Omega}^4 - \omega_{IBx}^2\omega^2\right) + \frac{\omega_{Pe}^2}{\Delta_{eB}}\left(\tilde{\omega}^4 - \omega_{eBx}^2\omega^2\right),$$

$$\varepsilon_{xy} = -\frac{\omega_{Pi}^2}{\Delta_{IB}}\left(i\tilde{\Omega}^2\omega_{IBz}\omega + \omega_{IBx}\omega_{IBy}\omega^2\right) + \frac{\omega_{Pe}^2}{\Delta_{eB}}\left(i\tilde{\omega}^2\omega_{eBz}\omega - \omega_{eBx}\omega_{eBy}\omega^2\right),$$

$$\varepsilon_{xz} = \frac{\omega_{Pi}^2}{\Delta_{IB}}\left(i\tilde{\Omega}^2\omega_{IBy}\omega - \omega_{IBx}\omega_{IBz}\omega^2\right) - \frac{\omega_{Pe}^2}{\Delta_{eB}}\left(i\tilde{\omega}^2\omega_{eBy}\omega + \omega_{eBx}\omega_{eBz}\omega^2\right),$$

$$\varepsilon_{yx} = \frac{\omega_{Pi}^2}{\Delta_{IB}}\left(i\tilde{\Omega}^2\omega_{IBz}\omega - \omega_{IBx}\omega_{IBy}\omega^2\right) - \frac{\omega_{Pe}^2}{\Delta_{eB}}\left(i\tilde{\omega}^2\omega_{eBz}\omega + \omega_{eBx}\omega_{eBy}\omega^2\right),$$

$$\varepsilon_{yy} = 1 + \frac{\omega_{Pi}^2}{\Delta_{IB}}\left(\tilde{\Omega}^4 - \omega_{IBy}^2\omega^2\right) + \frac{\omega_{Pe}^2}{\Delta_{eB}}\left(\tilde{\omega}^4 - \omega_{eBy}^2\omega^2\right), \quad (3)$$

$$\varepsilon_{yz} = -\frac{\omega_{Pi}^2}{\Delta_{IB}}\left(i\tilde{\Omega}^2\omega_{IBx}\omega + \omega_{IBy}\omega_{IBz}\omega^2\right) + \frac{\omega_{Pe}^2}{\Delta_{eB}}\left(i\tilde{\omega}^2\omega_{eBx}\omega - \omega_{eBy}\omega_{eBz}\omega^2\right),$$

$$\varepsilon_{zx} = -\frac{\omega_{Pi}^2}{\Delta_{IB}}\left(i\tilde{\Omega}^2\omega_{IBy}\omega + \omega_{IBx}\omega_{IBz}\omega^2\right) + \frac{\omega_{Pe}^2}{\Delta_{eB}}\left(i\tilde{\omega}^2\omega_{eBy}\omega - \omega_{eBx}\omega_{eBz}\omega^2\right),$$

$$\varepsilon_{zy} = \frac{\omega_{Pi}^2}{\Delta_{IB}}\left(i\tilde{\Omega}^2\omega_{IBx}\omega - \omega_{IBy}\omega_{IBz}\omega^2\right) - \frac{\omega_{Pe}^2}{\Delta_{eB}}\left(i\tilde{\omega}^2\omega_{eBx}\omega + \omega_{eBy}\omega_{eBz}\omega^2\right),$$

$$\varepsilon_{zz} = 1 + \frac{\omega_{Pi}^2}{\Delta_{IB}}\left(\tilde{\Omega}^4 - \omega_{IBz}^2\omega^2\right) + \frac{\omega_{Pe}^2}{\Delta_{eB}}\left(\tilde{\omega}^4 - \omega_{eBz}^2\omega^2\right),$$

where $\omega_{Pi}^2 = 4\pi e_{eff}^2 N_C / m_{eff}$ and $\omega_{Pe}^2 = 4\pi e^2 N_e / m$ are the ion and electron plasma frequencies, $\tilde{\omega}^2 = \omega_0^2 - \omega^2 - i\Gamma\omega$, $\tilde{\Omega}^2 = \Omega_\perp^2 - \omega^2 - i\Gamma\omega$, $\Delta_{eB} = \tilde{\omega}^2\left[\tilde{\omega}^4 - \left(\omega_{eBx}^2 + \omega_{eBy}^2 + \omega_{eBz}^2\right)\omega^2\right]$, $\Delta_{IB} = \tilde{\Omega}^2\left[\tilde{\Omega}^4 - \left(\omega_{IBx}^2 + \omega_{IBy}^2 + \omega_{IBz}^2\right)\omega^2\right]$.

We describe the dynamics of magnetic dipole moment of medium by Landau-Lifshitz dissipation equation [15]

$$\frac{\partial \mathbf{M}}{\partial t} = -\gamma \mathbf{M} \times \mathbf{H} - \mathbf{M}_R, \quad (4)$$

where $\mathbf{M}_R$ is the relaxation vector with components $M_{Rj} = \omega_{Rj}\left(M_j - \chi_0 H_j\right)$, $\omega_{Rj}$ is the frequency of relaxation along the axis $j = x, y, z$, $\gamma = ge/2mc$ is the gyromagnetic ratio, $\chi_0 = |M_0/H_0|$ is the static magnetic susceptibility, $M_0$ is the equilibrium magnetization, $\mathbf{H}$ is the magnetic field intensity in medium. The permeability of medium is defined at linear approximation for monochromatic field $H, M \sim e^{-i\omega t}$ by the expression for magnetic induction vector $B_i = \mu_{ij} H_j = H_i + 4\pi M_i$. The components of tensor $\mu_{ij}$ [11] are

$$\mu_{jx} = I_{jx} + \frac{4\pi\chi_0}{\Delta_H}\left(\omega_{Rx}\omega_{jx}^2 - \omega_{Hz}\omega_{jy}^2 + \omega_{Hy}\omega_{jz}^2\right),$$

$$\mu_{jy} = I_{jy} + \frac{4\pi\chi_0}{\Delta_H}\left(\omega_{Hz}\omega_{jx}^2 + \omega_{Ry}\omega_{jy}^2 - \omega_{Hx}\omega_{jz}^2\right), \quad (5)$$

$$\mu_{jz} = I_{jz} + \frac{4\pi\chi_0}{\Delta_H}\left(-\omega_{Hy}\omega_{jx}^2 + \omega_{Hx}\omega_{jy}^2 + \omega_{Rz}\omega_{jz}^2\right),$$

where $I_{ij}$ is the unit tensor, $\omega_{Hi} = \gamma H_{0i}$, $\omega_{xx}^2 = \omega_{Hx}^2 + (\omega_{Ry} - i\omega)(\omega_{Rz} - i\omega)$,
$\omega_{yx}^2 = \omega_{Hx}\omega_{Hy} + (\omega_{Rz} - i\omega)\omega_{Hz}$, $\omega_{xy}^2 = \omega_{Hx}\omega_{Hy} - (\omega_{Rz} - i\omega)\omega_{Hz}$, $\omega_{yy}^2 = \omega_{Hy}^2 + (\omega_{Rx} - i\omega)(\omega_{Rz} - i\omega)$,
$\omega_{xz}^2 = \omega_{Hx}\omega_{Hz} + (\omega_{Ry} - i\omega)\omega_{Hy}$, $\omega_{yz}^2 = \omega_{Hy}\omega_{Hz} - (\omega_{Rx} - i\omega)\omega_{Hx}$, $\omega_{zx}^2 = \omega_{Hx}\omega_{Hz} - (\omega_{Ry} - i\omega)\omega_{Hy}$,
$\omega_{zy}^2 = \omega_{Hy}\omega_{Hz} + (\omega_{Rx} - i\omega)\omega_{Hx}$, $\omega_{zz}^2 = \omega_{Hz}^2 + (\omega_{Rx} - i\omega)(\omega_{Ry} - i\omega)$.
$\Delta_H = (\omega_{Rx} - i\omega)[(\omega_{Ry} - i\omega)(\omega_{Rz} - i\omega) + \omega_{Hx}^2] + \omega_{Hy}[(\omega_{Ry} - i\omega)\omega_{Hy} - \omega_{Hx}\omega_{Hz}]$
$+ \omega_{Hz}[(\omega_{Rz} - i\omega)\omega_{Hz} - \omega_{Hx}\omega_{Hy}]$.

The electromagnetic field in bigyrotropic dielectric medium satisfies to the system of equations

$$(\nabla \times \mathbf{H})_i = \frac{\varepsilon_{ij}}{c}\frac{\partial E_j}{\partial t}, \quad (\nabla \times \mathbf{E})_i = \frac{\mu_{ij}}{c}\frac{\partial H_j}{\partial t}. \quad (6)$$

The system of equations (6) has solutions in the form of normal waves in bigyrotropic medium, i.e. the polariton waves.

### 3. Spectra of polaritons in bigyrotropic medium

From equations (6) we receive the system of equations for the monochromatic electric and magnetic fields $E, H \sim \exp(-i\omega t + ikz)$,

$$\frac{\omega}{c}\left(\varepsilon_{xx}E_x + \varepsilon_{xy}E_y + \varepsilon_{xz}E_z\right) - kH_y = 0, \qquad kE_y + \frac{\omega}{c}\left(\mu_{xx}H_x + \mu_{xy}H_y + \mu_{xz}H_z\right) = 0,$$

$$\frac{\omega}{c}\left(\varepsilon_{yx}E_x + \varepsilon_{yy}E_y + \varepsilon_{yz}E_z\right) + kH_x = 0, \quad -kE_x + \frac{\omega}{c}\left(\mu_{yx}H_x + \mu_{yy}H_y + \mu_{yz}H_z\right) = 0, \quad (7)$$

$$\varepsilon_{zx}E_x + \varepsilon_{zy}E_y + \varepsilon_{zz}E_z = 0, \qquad \mu_{zx}H_x + \mu_{zy}H_y + \mu_{zz}H_z = 0.$$

One can obtain the dispersion equation for polaritons in bigyrotropic dielectric medium at the presence of external magnetostatic field $\mathbf{H}_0$ by equating the determinant of equation system (7) to zero

$$\begin{vmatrix} \varepsilon_{xx} & \varepsilon_{xy} & \varepsilon_{xz} & 0 & -ck\omega^{-1} & 0 \\ \varepsilon_{yx} & \varepsilon_{yy} & \varepsilon_{yz} & ck\omega^{-1} & 0 & 0 \\ \varepsilon_{zx} & \varepsilon_{zy} & \varepsilon_{zz} & 0 & 0 & 0 \\ 0 & ck\omega^{-1} & 0 & \mu_{xx} & \mu_{xy} & \mu_{xz} \\ -ck\omega^{-1} & 0 & 0 & \mu_{yx} & \mu_{yy} & \mu_{yz} \\ 0 & 0 & 0 & \mu_{zx} & \mu_{zy} & \mu_{zz} \end{vmatrix} = 0, \quad (8)$$

We can get from equation (8) the squared refractive index of medium $n^2 = \varepsilon_{ij}\mu_{ij}$ which coincides with $n$ having obtained in the case of two axes dispersion equation [6].

Let us consider the spectrum of polaritons in bigyrotropic medium when the magnetostatic field is perpendicular to the wavevector $H_{0x} = H_{0\perp}, H_{0y} = 0, H_{0z} = 0$. In this case the tensors of permittivity and permeability have the components

$$\varepsilon_{xx} = \tilde{\varepsilon}, \quad \varepsilon_{yy} = \varepsilon_{zz} = \varepsilon_d, \quad \varepsilon_{yz} = -i\varepsilon_{nd}, \quad \varepsilon_{zy} = i\varepsilon_{nd}, \quad \varepsilon_{xy} = \varepsilon_{yx} = \varepsilon_{xz} = \varepsilon_{zx} = 0,$$

$$\mu_{xx} = \tilde{\mu}, \quad \mu_{yy} = \mu_{zz} = \mu_d, \quad \mu_{yz} = -\mu_{zy} = -i\mu_{nd}, \quad \mu_{xy} = \mu_{xz} = \mu_{yx} = \mu_{zx} = 0,$$

where $\tilde{\varepsilon} = 1 + \dfrac{\omega_{Pi}^2}{\tilde{\Omega}^2} + \dfrac{\omega_{Pe}^2}{\tilde{\omega}^2}$, $\varepsilon_d = 1 + \dfrac{\omega_{Pi}^2 \tilde{\Omega}^2}{\tilde{\Omega}^4 - \omega_{IBj}^2 \omega^2} + \dfrac{\omega_{Pe}^2 \tilde{\omega}^2}{\tilde{\omega}^4 - \omega_{eBj}^2 \omega^2}$, $\varepsilon_{nd} = \dfrac{\omega_{Pi}^2 \omega_{IBx} \omega}{\tilde{\Omega}^4 - \omega_{IBj}^2 \omega^2} + \dfrac{\omega_{Pe}^2 \omega_{eBx} \omega}{\tilde{\omega}^4 - \omega_{eBj}^2 \omega^2}$

$$\tilde{\mu} = 1 + 4\pi\chi_0 \frac{i\omega_R}{\omega + i\omega_R}, \quad \mu_d = 1 + 4\pi\chi_0 \frac{\omega_R^2 + \omega_{Hj}^2 - i\omega_R\omega}{\omega_{Hj}^2 + \omega_R^2 - \omega^2 - i2\omega_R\omega}, \quad \mu_{nd} = \frac{4\pi\chi_0 \omega_{Hj}\omega}{\omega_{Hj}^2 + \omega_R^2 - \omega^2 - i2\omega_R\omega},$$

$\omega_{Rx} = \omega_{Ry} = \omega_{Rz} = \omega_R$, and the dispersion equation (8) gains the form

$$\frac{\omega^4}{c^4}\tilde{\varepsilon}\tilde{\mu}(\varepsilon_d^2 - \varepsilon_{nd}^2)(\mu_d^2 - \mu_{nd}^2) - \frac{\omega^2}{c^2}k^2\left[\tilde{\varepsilon}\varepsilon_d(\mu_d^2 - \mu_{nd}^2) + \tilde{\mu}\mu_d(\varepsilon_d^2 - \varepsilon_{nd}^2)\right] + k^4\varepsilon_d\mu_d = 0. \quad (9)$$

In the case when the magnetostatic field is parallel to the wavevector $H_{0x} = 0, H_{0y} = 0, H_{0z} = H_{0\parallel}$ the tensors of permittivity and permeability have the components

$$\varepsilon_{xx} = \varepsilon_{yy} = \varepsilon_d, \quad \varepsilon_{zz} = \tilde{\varepsilon}, \quad \varepsilon_{yx} = -\varepsilon_{xy} = i\varepsilon_{nd}, \quad \varepsilon_{xz} = \varepsilon_{yz} = \varepsilon_{zx} = \varepsilon_{zy} = 0,$$

$$\mu_{xx} = \mu_{yy} = \mu_d, \quad \mu_{zz} = \tilde{\mu}, \quad \mu_{xy} = -\mu_{yx} = -i\mu_{nd}, \quad \mu_{xz} = \mu_{yz} = \mu_{zx} = \mu_{zy} = 0,$$

and the dispersion equation (8) acquires the form

$$\frac{\omega^4}{c^4}(\varepsilon_d^2 - \varepsilon_{nd}^2)(\mu_d^2 - \mu_{nd}^2) - 2\frac{\omega^2}{c^2}k^2(\varepsilon_d\mu_d - \varepsilon_{nd}\mu_{nd}) + k^4 = 0. \quad (10)$$

The polariton spectra (9) and (10) are presented in Fig. 1a at $H_{0\perp}$ and in Fig. 1b at $H_{0//}$. In these Figures the positive real parts $\operatorname{Re}(\omega_i)$ of frequencies are shown.

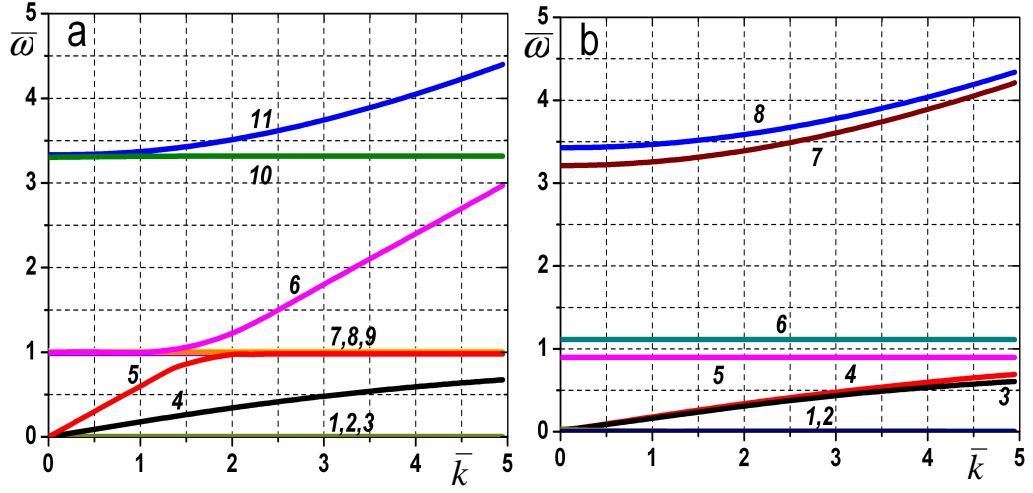

Fig. 1. The polariton spectra in bigyrotropic medium: a) at $H_{0\perp}$ and b) at $H_{0\|}$; $M_0 = 100\,G$, $\chi_0 = 4$, $\Gamma = 10^4\,s^{-1}$, $\omega_R/\omega_M = 0.1$, $\omega_M = 4\pi\gamma M_0$. The frequency and wavevector are presented in the dimensionless units $\bar{\omega} = \omega/\Omega_\perp$, $\bar{k} = ck/\Omega_\perp$.

For $H_{0\perp}$ (Fig. 1a) the branches 5 and 6 of polariton spectrum in bigyrotropic medium coincide with branches of polariton spectrum having obtained in the paper [4]. The branches 1, 2, 3, 4, 7, 8, 9, 10, 11 appear in the spectrum due to the presence of components in tensors $\varepsilon, \mu$ which describe the bigyrotropical properties of medium. In this case a gap between branches 6 and 7 takes place. The polariton spectrum at $H_{0\|}$ (Fig. 1b) is characterized by less number of branches then spectrum at $H_{0\perp}$ and the presence of two gaps (between branches 5 and 6, and between 6 and 7). The position and width of gaps depend on the direction and intensity of the external magnetostatic field (see Fig. 1a, Fig. 1b). Thus, the number and form of polariton spectrum branches in bigyrotropic medium essentially depend on direction and intensity of the magnetostatic field. One can control the polariton spectrum in the medium by changing the intensity and direction of the magnetostatic field.

The natural waves (polariton waves) are excited in the medium with determined wavevectors by the electromagnetic field with the given frequency $\omega$. We obtain the solutions for wavevectors of polariton waves from the dispersion equations (9) and (10) presented as $k^4 - 2\bar{a}_{1,2}k^2 + \bar{b}_{1,2} = 0$,

$$k_{1,2}^{\pm} = \frac{\omega}{c} n_{1,2}^{\pm}, \qquad (11)$$

where $n_{1,2}^{\pm} = \left[\bar{a}_{1,2} \pm \left(\bar{a}_{1,2}^2 - \bar{b}_{1,2}\right)^{1/2}\right]^{1/2}$ is the refractive index of medium, $\bar{a}_1 = \frac{1}{2}\left[\frac{\tilde{\varepsilon}}{\mu_d}\left(\mu_d^2 - \mu_{nd}^2\right) + \frac{\tilde{\mu}}{\varepsilon_d}\left(\varepsilon_d^2 - \varepsilon_{nd}^2\right)\right]$, $\bar{b}_1 = \frac{\tilde{\varepsilon}\tilde{\mu}}{\varepsilon_d \mu_d}\left(\varepsilon_d^2 - \varepsilon_{nd}^2\right)\left(\mu_d^2 - \mu_{nd}^2\right)$ at $\mathbf{H}_0 \perp \mathbf{k}_z$; and $\bar{a}_2 = \varepsilon_d \mu_d - \varepsilon_{nd}\mu_{nd}$, $\bar{b}_2 = \left(\varepsilon_d^2 - \varepsilon_{nd}^2\right)\left(\mu_d^2 - \mu_{nd}^2\right)$ at $\mathbf{H}_0 \parallel \mathbf{k}_z$. One can see from expression (11) that the birefringence $n_{1,2}^{\pm}$ takes place at the both configurations of fields $\mathbf{H}_0 \perp \mathbf{k}_z$ and $\mathbf{H}_0 \parallel \mathbf{k}_z$.

### 4. Velocity of polaritons

The velocity of polaritons in bigyrotropic medium depends on the frequency of electromagnetic field, on intensity and direction of the external magnetostatic field. We can obtain the velocity of polariton by substituting the wavevectors (11) in expression $v_i = \mathrm{Re}(dk_i/d\omega)^{-1}$ depending on the value of refractive index

$$v_{1,2}^{\pm} = c\left[\mathrm{Re}\left(n_{1,2}^{\pm} + \omega\, dn_{1,2}^{\pm}/d\omega\right)\right]^{-1}. \qquad (12)$$

The dependence on frequency $\omega$ of polariton velocity in bigyrotropic medium at zero intensity of the magnetostatic field $H_0 = 0$ is presented in Fig. 2.

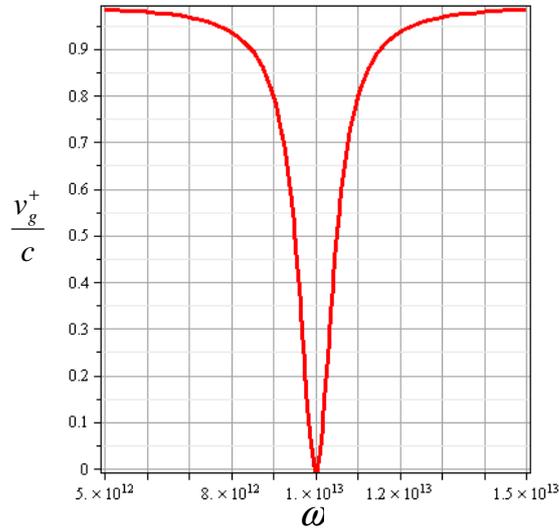

Fig. 2. The dependence on frequency $\omega$ of normalized polariton velocity $v_g^+/c$

in bigyrotropic medium at $H_0 = 0$.

The polariton velocities $v_g^+$ and $v_g^-$ at $H_0 = 0$ coincide for all frequencies. The velocities of low-frequency and high-frequency polaritons at $H_0 = 0$ tend to speed of light in vacuum far from resonance frequencies of medium.

The dependence on directions of magnetostatic field $H_{0\perp}$ and $H_{0\|}$ of polariton velocities is presented in Fig. 3 for the field intensity $H_0 = 2500\,Oe$. We choose the parameters of bigyrotropic medium: $\omega_{Pi} = 10^{12}\,s^{-1}$, $\omega_{Pe} = 10^{16}\,s^{-1}$, $\Omega_\perp = 10^{13}\,s^{-1}$, $\omega_0 = 10^{17}\,s^{-1}$, $\chi_0 = 4$, $\Gamma = 10^4\,s^{-1}$, $\omega_R = 3\cdot 10^9\,s^{-1}$.

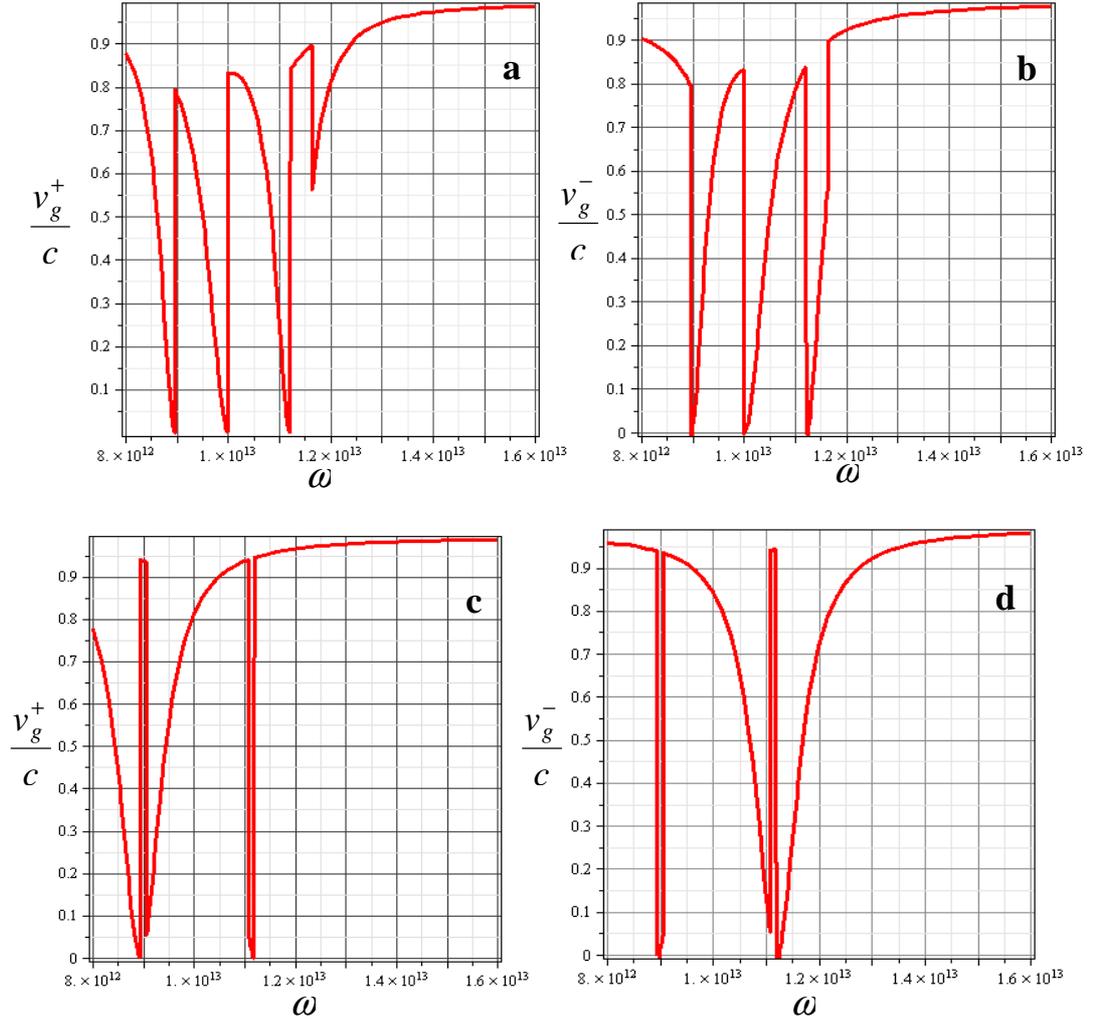

Fig. 3. The dependence on frequency $\omega$ of polariton velocities $v_g^\pm$ for field configurations in bigyrotropic medium: a) and b) at $H_{0\perp}$; c) and d) at $H_{0\|}$. The intensity of external magnetostatic field is $H_0 = 2500\,Oe$.

From the analysis of polariton velocity dependence on the frequency, direction and intensity of external magnetostatic field in bigyrotropic medium with the given parameters we can see that close to the frequency of lattice resonance $\Omega_\perp$ the polariton velocity tends to zero (Fig. 3). It is caused by the presence of several gaps in the polariton spectrum. The magnetostatic field $\mathbf{H}_0$ increases the number of resonance frequencies nearby the lattice resonance frequency $\Omega_\perp$ that leads to the growth of number of gaps in the spectrum where polariton velocity is equal zero. The dispersion dependences of polariton velocities $v_g^+(\omega)$ and $v_g^-(\omega)$ essentially depend on the direction of the magnetostatic field $\mathbf{H}_0$ at its fixed intensity. The changing of the magnetostatic field from direction $H_{0\perp}$ to direction $H_{0\parallel}$ relatively to the wavevector of polariton wave shifts the central resonance dip to low frequency range for polaritons with velocity $v_g^+$ and to high frequency range for polaritons with velocity $v_g^-$.

## 5. Conclusion

The polariton spectra in bigyrotropic medium essentially depend on the direction relative to the wavevector and intensity of the external magnetostatic field. Thus the changing of direction or intensity of magnetostatic field allows controlling the polariton spectrum and polariton velocity. Based on described properties of polariton spectrum in bigyrotropic medium one can design the components of the terahertz and optical data-transmission line [16, 17]: the controllable filters, delay lines, logic gates, photonic crystal, etc.

For example, the delay line for the electromagnetic radiation with the frequency $\omega = 1.1 \cdot 10^{13} s^{-1}$ and intensity of external magnetostatic field $H_0 = 2500\, Oe$ allows decreasing the polariton velocity $v_g^+$ from $2.85 \cdot 10^{10} cm/s$ to $0.6 \cdot 10^{10} cm/s$ by changing direction of magnetostatic field at $90^0$ for the crystal with the next parameters: $\Omega_\perp = 10^{13} s^{-1}, \omega_0 = 10^{17} s^{-1}$, $\chi_0 = 4, \Gamma = 10^4 s^{-1}, \omega_R = 3 \cdot 10^9 s^{-1}$. In this case the signal delay time at centimeters of length of the crystal $t = 1/v_g^+$ will vary from $0.35 \cdot 10^{-10} s$ to $1.67 \cdot 10^{-10} s$, so it increases to five times.